\documentclass{article}

\usepackage{microtype}
\usepackage{graphicx}
\usepackage{subcaption}
\usepackage{booktabs} 
\usepackage{xurl}

\usepackage{hyperref}

\usepackage[accepted]{icml2024}

\usepackage{amsmath}
\usepackage{amssymb}
\usepackage{mathtools}
\usepackage{amsthm}

\usepackage[capitalize,noabbrev]{cleveref}

\theoremstyle{plain}

\theoremstyle{definition}

\theoremstyle{remark}

\usepackage[textsize=tiny]{todonotes}

\icmltitlerunning{AstroPT: Scaling Large Observation Models for Astronomy}

\begin{document}

\twocolumn[
\icmltitle{AstroPT: Scaling Large Observation Models for Astronomy}



\icmlsetsymbol{equal}{*}

\begin{icmlauthorlist}
\icmlauthor{Michael J. Smith}{aspia,iac,utbd}
\icmlauthor{Ryan J. Roberts}{aspia,ljmu}
\icmlauthor{Eirini Angeloudi}{iac,ull}
\icmlauthor{Marc Huertas-Company}{iac,ull,psl,upc}
\end{icmlauthorlist}

\icmlaffiliation{aspia}{Aspia Space}
\icmlaffiliation{iac}{Instituto de Astrof\'isica de Canarias (IAC)}
\icmlaffiliation{ull}{Departamento Astrof\'isica, Universidad de la Laguna}
\icmlaffiliation{ljmu}{Astrophysics Research Institute, Liverpool John Moores University}
\icmlaffiliation{utbd}{UniverseTBD}
\icmlaffiliation{psl}{Observatoire de Paris, LERMA, PSL University}
\icmlaffiliation{upc}{Universit\'e Paris-Cit\'e}

\icmlcorrespondingauthor{Michael J. Smith}{mike@mjjsmith.com}

\icmlkeywords{Machine Learning, ICML}

\vskip 0.3in
]



\printAffiliationsAndNotice{}  

\begin{abstract}
    This work presents AstroPT, an autoregressive pretrained transformer developed with astronomical use-cases in mind. The AstroPT models presented here have been pretrained on 8.6 million $512 \times 512$ pixel {\it grz}-band galaxy postage stamp observations from the DESI Legacy Survey DR8. We train a selection of foundation models of increasing size from 1 million to 2.1 billion parameters, and find that AstroPT follows a similar saturating log-log scaling law to textual models. We also find that the models' performances on downstream tasks as measured by linear probing improves with model size up to the model parameter saturation point. We believe that collaborative community development paves the best route towards realising an open source `Large Observation Model'---a model trained on data taken from the observational sciences at the scale seen in natural language processing. To this end, we release the source code, weights, and dataset for AstroPT under the MIT license, and invite potential collaborators to join us in collectively building and researching these models.
\end{abstract}

\section{On `Large Observation Models'}
\label{introduction}

We find ourselves in a new era of connectionism. This era is dominated by large language models trained on
web-scale data, with a rapid parameter growth fertilised by the discovery of
predictable `neural scaling laws'---power laws that can be used to estimate a
model's performance given its parameter count and training set size
\citep{ref_cortes1993,ref_kaplan2020}.
In the language domain, the law that currently best describes model scaling was
introduced in \citet{ref_hoffmann2022}. This `Chinchilla' scaling law
established that for every additional neural parameter added to the network,
around twenty additional textual data tokens must also be added to train a
model in a training compute optimal regime.  It is also often wise to
`overtrain' a model---or train at a token to parameter ratio larger than
20:1---which leads to a more performant model at a lower
parameter count and therefore reduces compute cost at inference time. Overtraining is
both environmentally and economically beneficial if a model's total lifetime
compute costs are inference heavy
\citep{ref_touvron2023,ref_touvron2023b,ref_zhang2024}.  The discovery of the
Chinchilla scaling law, and the need to minimise compute cost at inference time
means that we have nearly exhausted high-quality publicly available reserves of
textual data for training large neural models \citep{ref_sevilla2022,ref_xue2023}.  Some potential
solutions to this `token crisis' have already been investigated in the
literature: for example, one can repeat training dataset epochs multiple times
without significant performance degradation \citep{ref_xue2023}, or one can
also turn to the use of synthetic data to generate tokens at scale
\citep{ref_silver2018,ref_gunasekar2023}. We believe that multimodality can
provide a further solution.

Large autoregressive models can process and digest tokens of different
modalities \citep{ref_reed2022gato}. We can therefore build a model that is
capable of learning from modalities that have an abundance of tokens, an abundance
that is particularly notable in the observational sciences
\citep{ref_smith2023,ref_smith2023earthpt}. 
To give two representative examples, in astronomy the Vera Rubin
Observatory (VRO) will observe over twelve billion $16
\times 16$ pixel patch `tokens' per night when conducting LSST
\citep{ref_lsst}, and in earth observation ESA's Sentinel-2 mission produces
over six trillion land-observation tokens on a five day global revisit cadence.
There are of course many more missions in the observational sciences besides
VRO's LSST and ESA's Copernicus, and a compilation of all useful observational
data would certainly dwarf any natural textual dataset in volume. Such a dataset
could be combined with aligned textual or other observational data and used to
train a foundation model---and so, as large neural models can successfully connect
concepts across modalities \citep[e.g.][]{ref_aghajanyan2023}, unlocking these
data would go some way towards solving the token crisis.

We will leave true cross-modal training to future work and concentrate here
on using astronomical data as the first step towards the goal of training a
general `Large Observation Model' (LOM).  Before we dive into describing our
model, we will briefly summarise the field as it stands to provide further motivation
for our specific approach.  For the purposes of this
literature review let us consider a `foundation model' as a model that
comprises of two training steps. The first step being a computationally
expensive unsupervised `pretraining' routine, and the second step being a
relatively computationally cheap supervised finetuning or conditional prompting
routine. If we view the field through this lens, two
main avenues\footnote{
    There are of course other interesting approaches that do not fit into this
    neat dichotomy, for example \citet{ref_charnock2018, ref_jeffrey2021,
    ref_walmsley2022reps, ref_walmsley2024}.
} come into focus---that is,
contrastive learning and generative modelling. Let us discuss these techniques
in the subsections below\footnote{
    There are many more applications of contrastive learning and generative
    modelling in astronomy---far more than we can include in this paper.  We
    therefore direct the interested reader to \citet{ref_huertas2023} for
    further reading on contrastive learning, and to \S\S6--8 of
    \citet{ref_smith2023} for generative modelling.
}.

\subsection{Contrastive learning}

The first contrastive self-supervised neural network used in astronomy, and
perhaps the first work that could be considered an `astronomical foundation
model' in the modern sense is `SkyNet' \citep{ref_graff2014}. SkyNet's
pretraining routine is driven by contrastive divergence learning, where a
model is trained by minimising the difference between the positive pairs of
training set examples and reconstructions
\citep{ref_hinton2002,ref_hinton2006}.  Once SkyNet is trained,
\citet{ref_graff2014} found that it could be finetuned for downstream
astronomical tasks.

We can derive positive and negative pairs directly from
training data. \citet{ref_hayat2021} do precisely this and pretrain a SimCLR
model on {\it ugriz}-band galaxy observations \citep{ref_chen2020}.
Along similar lines, \citet{ref_sarmiento2021} pretrain a SimCLR on MaNGA galaxy data cubes
\citep{ref_bundy2014}. \citet{ref_slijepcevic2022} show that the BYOL model
can produce meaningful embeddings when trained on only positive radio galaxy
pairs \citep{ref_grill2020}, and \citet{ref_akhmetzhanova2024} show that the 
VICReg contrastive framework can learn useful embeddings from cosmological
simulations \citep{ref_bardes2021}.

There have also been cross-modal contrastive approaches to representation
learning in astronomy. \citet{ref_lanusse2023} describe using a CLIP-like
method \citep{ref_radford2021} to align representations of galaxy images
and their spectra.  \citet{ref_mishra2024} take a similar approach with their
PaperCLIP model.  They align textual information and astronomical imagery by
training a CLIP criterion on imagery-text pairs derived from telescope
observation proposals.

From these studies, we can gather that contrastive learning works across
multiple domains even when positive pairs are sourced from different
instruments and modalities. The one hindrance to scaling contrastive learning
approaches across and within modalities is the need to generate these positive
pairs. To do so we either need to impose our domain knowledge onto the data
pair generation routine, or otherwise labouriously crossmatch intermodal pairs.

\subsection{Generative modelling}

We can also use generative modelling to create scientifically useful compressed
representations of astronomical objects: \citet{ref_schawinski2018} pretrained 
a variational autoencoder \citep[VAE;][]{ref_kingma2013,ref_lample2017} on the
task of galaxy image recreation. Their model was able to learn
unentangled parameters that described physical properties of the galaxies, and
furthermore was able to simulate a realistic galaxy given a set of learnt
properties.  A similar approach was taken in \citet{ref_spindler2020}, who
trained a VAE on the pretraining task of recreating galaxy images. They
found that their `AstroVaDEr' model was capable of learning useful embeddings
and of restoring galaxy imagery to a high degree of accuracy.
\citet{ref_smith2022} trained a diffusion model on the pretraining task of
replicating galaxy images \citep{ref_ho2020}.  Although the resulting diffusion
model was not intended to serve as a foundation for downstream tasks,
\citet{ref_karchev2022} were able to take the pretrained model and 
use it to perform the out of distribution task of reversing gravitational
lensing events, without any further training. 
Non-textual autoregressive generative approaches have found use across astronomical domains.
For instance, \citet{ref_zanisi2021} show that an autoregressive causally-masked PixelCNN++
model \citep{ref_oord2016,ref_oord2016b,ref_salimans2017} is capable of
quantifying the morphological differences between real and simulated galaxy
observations.
Similarly, \citet{ref_muthukrishna2021} show that an autoregressive temporal
convolutional neural network is capable of learning embeddings that can then
be used to detect outliers in their dataset---in this case stellar transients.
While all of the above approaches light viable paths towards building our
foundation model, we believe that decoding transformer models pretrained in a
causal autoregressive manner are currently the most promising approach to
realising a sizable LOM that has open code and weights, and that has been
pretrained on openly available data. This is largely due to non-technical
sociological factors, and we will discuss our reasoning in detail in
\S\ref{sec_model}.

Causally-masked language foundation models have also been explored in astronomy.
\citet{ref_nguyen2023} presented AstroLLaMA---a LLaMA-2-7B model
\citep{ref_touvron2023b} that was finetuned on high-quality astronomical
research text. \citet{ref_perkowski2024} extended AstroLLaMA into a
conversational chatbot by further finetuning on a mix of synthetic
and human-generated astronomy question-answer pairs. 
Non-textual transformer-based LOMs have been explored in the literature too.
As just one example, \citet{ref_leung2023} describe an encoder-decoder
transformer-based model for stellar information extraction \citep{ref_aiayn}.
While a robust and innovative approach, \citet{ref_leung2023} leave some open
questions which we hope to complement with this work: that is, can we scale
neural networks on astronomical observation data just as we have done in the
textual domain, and do we need the computational and architectural overhead of
pretraining a full encoder-decoder transformer architecture to teach our models
scientifically useful information?

This manuscript is structured as follows.  In the next section we will describe
why we believe that a causal transformer is currently the most promising
architectural candidate for building our astronomical foundation model
(\S\ref{sec_model}), and the dataset we used to train our candidate model
(\S\ref{sec_dataset}). In \S\ref{sec_results} we present our results and
discussion.  Finally, we bring this paper to an end with some closing remarks
in \S\ref{sec_conclusion}.

\section{Methods}

Here we describe the hyperparameters and training routine of AstroPT, and the
dataset we used to train the model.

\subsection{AstroPT}
\label{sec_model}

Decoding transformer architectures \citep{ref_aiayn,ref_radford2018gpt1} have a
number of benefits that make them well suited as architectural candidates for
training LOMs.  Firstly, decoding transformers are
efficient at pretraining time due to their causal self-attention mask
which ensures that every item in an input sequence creates a signal to be
backpropagated.  
Secondly, the decoding transformer's ubiquity within the
open source community has lead to an active `bazaar' of enthusiasts and
researchers providing innovations and contributions to the general architecture
\citep{ref_phang2022,ref_touvron2023,ref_liu2023}. 
As the resource required to develop these large models is substantial, it would be
beneficial for the field if the astronomical and other observational sciences
follow and contribute back to the upstream community's work as much as possible
when developing application-driven machine learning models
\citep{ref_raymond1999,ref_smith2023,ref_rolnick2024}.
Also, we can follow Sutton's \yrcite{ref_sutton2019} `Bitter Lesson' to a logical
endpoint and arrive at the conclusion that the specific neural
architecture used to learn a pretraining task does not matter so long as it
scales efficiently and is appropriate for the task at hand \citep[see for
example][]{ref_peng2023,ref_peng2024,ref_bachmann2023,ref_smith2023google,ref_huh2024}.  
Taking a generative approach to embedding extraction removes
the need for us to impose external knowledge on the network, something that
is required to create physically-consistent positive and negative pairs in
a contrastive learning regime.
And an autoregressive training process allows us to incorporate multiple
modalities via simple token chaining, which is not possible with other
contrastive or generative approaches.  For these reasons it is prudent to build
foundation models off of the task-appropriate general neural architecture that
has seen the most development time and that enjoys an active community, and to
train the general architecture within an autoregressive generative regime.
These criteria are fulfilled by the decoding
causally masked transformer model.

\begin{figure}[htb]
\vskip 0.2in
\begin{center}
\centerline{\includegraphics[width=0.6\columnwidth]{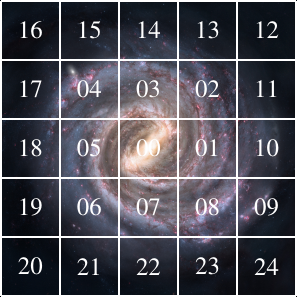}}
\caption{
    In this work we train AstroPT on the surrogate task of predicting the next token in a
    `spiralised' sequence of galaxy image patches. The above image shows the
    token feed order. As the galaxies are in the centre of each postage stamp,
    this set up allows us to seamlessly pretrain and run inference on
    differently sized galaxy postage stamps.
}
\label{fig_galaxyim}
\end{center}
\vskip -0.2in
\end{figure}

With these arguments in mind we propose AstroPT---the Astronomical Pretrained
Transformer. AstroPT is a GPT-2 like model that has been repurposed for
regression tasks \citep{ref_radford2019gpt2,ref_smith2023earthpt}. To this end
we replace the textual tokeniser with a multilayer perceptron (MLP) tokeniser that
feeds directly into AstroPT, and merge position embeddings with the tokens in
the standard additive way. In this way, AstroPT is capable of learning any
general autoregressive task. In this work we train on a dataset of galaxy
imagery, which we describe in the next subsection. To tokenise our
galaxy dataset we follow \citet{ref_dosovitskiy2020vit} and define a token as
a $16\times16$ pixel patch, and we follow \citet{ref_he2022} and
\citet{ref_elnouby2024} by independently standardising each $16 \times 16$
pixel galaxy image patch to have a mean of zero and a standard deviation of one
before passing them into AstroPT's MLP tokeniser.  We define the learning
objective for AstroPT as predicting the next token in a sequence under the
Huber loss criterion \citep{ref_huber1964}. We feed the tokens into AstroPT in
a `spiral' order, as shown in Fig.~\ref{fig_galaxyim}.

We train a selection of models from 1 million to 2.1 billion parameters,
following closely the hyperparameters chosen in \citet{ref_biderman2023} for
comparison's sake. We document our hyperparameters in Tab.~\ref{tab_hparam}.

\begin{table*}[htb]
\caption{
    Chosen hyperparameters for our AstroPT models. The discrepancy in the
    number of model parameters between the comparable language models and
    AstroPT models is due to differences in number of parameters between the
    models' respective tokenisers. The Pythia family of models are described in
    \citet{ref_biderman2023}, the GPT-Neo-125M model is described in
    \citet{ref_black2021}, and the OPT-125M model is described in
    \citet{ref_zhang2022}.  
}
\label{tab_hparam}
\vskip 0.15in
\begin{center}
\begin{small}
\begin{sc}
\begin{tabular}{lccccr}
\toprule
$N$ params. & Layers & Model Dim & Heads & Learning Rate & Comparable Language Models \\
\midrule
1M          & 4 & 128 & 8 & $10 \times 10^{-4}$ & ---\\
5M          & 6 & 256 & 8 & $10 \times 10^{-4}$ & ---\\
12M         & 6 & 384 & 8 & $10 \times 10^{-4}$ & ---\\
21M         & 6 & 512 & 8 & $10 \times 10^{-4}$ & Pythia-70M\\
89M         & 12 & 768 & 12 & $6 \times 10^{-4}$ & Pythia-160M, GPT-Neo-125M, OPT-125M\\
309M        & 24 & 1024 & 16 & $3 \times 10^{-4}$ & Pythia-160M\\
830M        & 16 & 2048 & 8 & $3 \times 10^{-4}$ & Pythia-1.0B\\
2.1B        & 26 & 2560 & 32 & $1.6 \times 10^{-4}$ & ---\\
\bottomrule
\end{tabular}
\end{sc}
\end{small}
\end{center}
\vskip -0.1in
\end{table*}

\subsection{Dataset and data processing}
\label{sec_dataset}

Our dataset comprises of 8.6 million Dark Energy Spectroscopic Instrument
Legacy Survey (DESI-LS) data release 8 galaxies
\citep{ref_dey2019,ref_walmsley2023}. 
DESI-LS is an extensive sky survey comprising three complementary public projects (DECaLS, BASS, and MzLS) aimed at imaging approximately 14\,000 deg$^2$ of the extragalactic sky from both the northern and southern hemispheres, using telescopes at the Kitt Peak National Observatory and the Cerro Tololo Inter-American Observatory.  
DESI-LS also includes imaging from the Dark Energy Survey (DES), which, although not part of the core DESI-LS, uses the same instrumentation as DECaLS and contributes an additional 5\,000 deg$^2$ of {\it g,r,z} imaging.
The galaxy catalogue used in this work comprises of all extended sources within
the DESI-LS source database that have an {\it r}-band magnitude greater than
19, and a surface brightness less than $18\;\text{mag}\,\text{arcsec}^{-2}$ (to
avoid stellar contamination).  We use this catalogue to build a dataset of
source-centred {\it grz}-band postage stamps of shape $512 \times 512$ in JPEG
format, at a resolution of $0.262\;\text{arcsec}\,\text{pixel}^{-1}$.  Each
galaxy is crossmatched with a set of emergent physical properties from the NSA
\citep{ref_aguado2019}, OSSY Type 1 AGN catalogue \citep{ref_oh2015}, Arecibo
Legacy Fast ALFA survey \citep{ref_haynes2018}, the MPA-JHU SDSS-DR7 derived
properties catalogue \citep{ref_abazajian2009}, the DESI photometric redshift
catalogue \citep{ref_zhou2021}, and Galaxy Zoo morphological classification
labels \citep{ref_walmsley2023}.  We can use these crossmatched properties to
validate our models' acquired physical knowledge.  We downloaded the data
directly from DESI-LS using their API, and have reuploaded the resulting galaxy
postage stamps and crossmatched property metadata to HuggingFace. We split our
galaxy dataset into three sets, a training set that contains 98\% (or
8\,480\,000)  of our images, and a test and validation set that each contain
1\% (or 86\,500) of our images.  We pretrain AstroPT on our training set, and
finetune our downstream tasks on the validation set, with our downstream task
results inferred from the test set.

\section{Results and discussion}
\label{sec_results}

\begin{figure}[htb]
\vskip 0.2in
\begin{center}
\centerline{\includegraphics{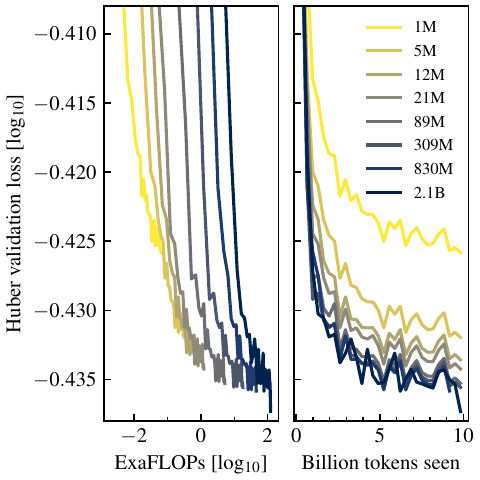}}
\caption{
    Validation set losses over our full training runs. The left plot shows
    the validation loss per training floating point operation (FLOP), and the right plot shows
    the validation loss per $16 \times 16$ image patch token seen.  Each run is
    labelled with the total neural parameter count as crossmatched in
    Tab.~\ref{tab_hparam}.
}
\label{fig_scaling}
\end{center}
\vskip -0.2in
\end{figure}

\begin{figure*}[htbp]
\begin{subfigure}{\textwidth}
\centering
    \centerline{\includegraphics{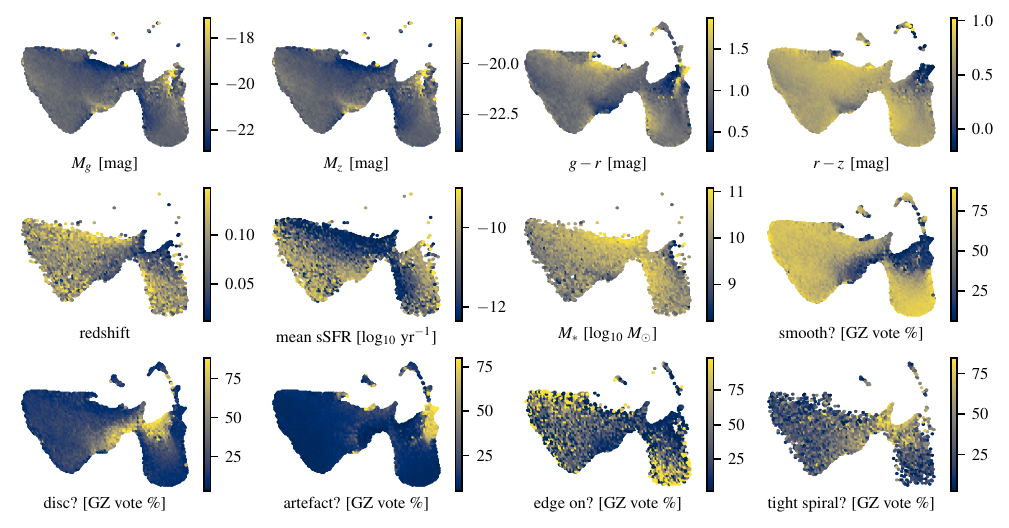}}
    \caption{
        Here we show our UMAP projected two dimensional embedding plots for AstroPT-1M.
    }
\end{subfigure}
\begin{subfigure}{\textwidth}
\vskip 0.1in
\centering
    \centerline{\includegraphics{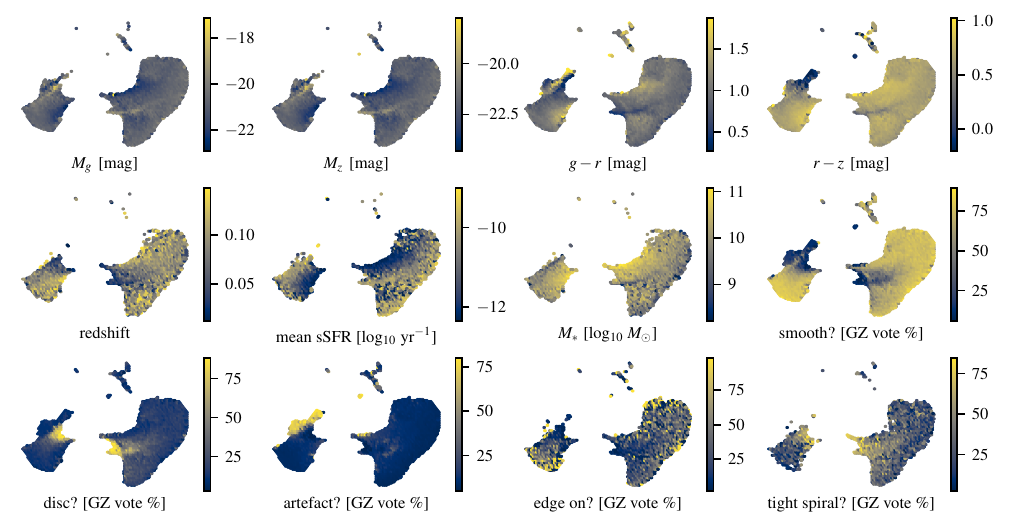}}
    \caption{
        Here we show our UMAP projected two dimensional embedding plots for AstroPT-89M.
    }
\end{subfigure}
\caption{
    Results from our AstroPT-1M embedding UMAP projections (upper), and
    AstroPT-89M embedding UMAP projections (lower).
    We colour the hex bins in both plots with a selected set of emergent physical properties of the galaxies. 
    We find significant structure, signifying that the model has learned physically meaningful representations of the dataset.
    In the above plots `$M_g$' and `$M_z$' are the absolute magnitudes in the $g$ and $z$ bands, `mean sSFR' is the mean specific star formation rate, and `$M_*$' is the stellar mass.  `smooth?', `disc?', `artefact?', `edge on?' and `tight spiral?' are Galaxy Zoo survey responses for these morphological features.
    Our metadata sources are described further in \S\ref{sec_dataset}.}
\label{fig_embeddings}
\end{figure*}

\begin{figure*}[htb]
\vskip 0.2in
\begin{center}
    \centerline{\includegraphics{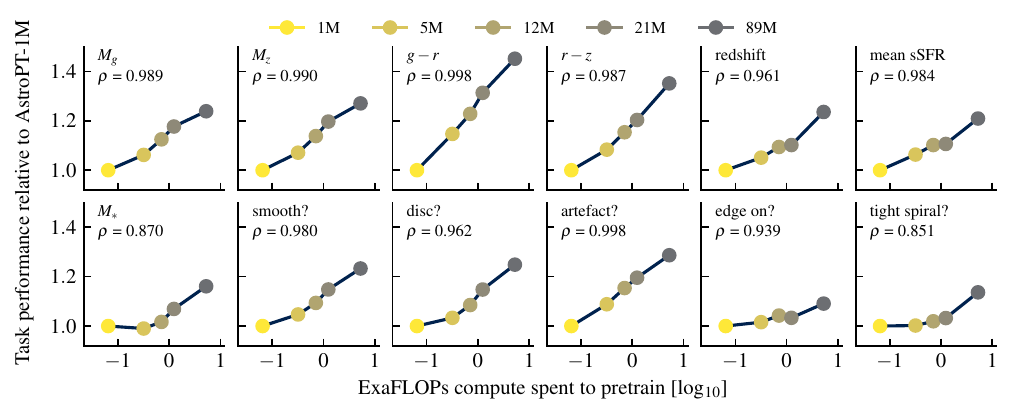}}
    \caption{
        Here we show our relative linear probe performances per pretraining FLOP spent on a selection of scientifically-meaningful downstream tasks. The markers are coloured according to the models' parameter counts.
        We run a Spearman's $\rho$ fit and find in all cases a strong positive correlation between downstream task performance and model size, meaning that a larger model has more informative embeddings.
        In this plot `$M_g$' and `$M_z$' are the absolute magnitudes in the $g$ and $z$ bands, `mean sSFR' is the mean specific star formation rate, and `$M_*$' is the stellar mass.  `smooth?', `disc?', `artefact?', `edge on?' and `tight spiral?' are Galaxy Zoo survey responses for these morphological features.
    Our metadata sources are described further in \S\ref{sec_dataset}.
    }
\label{fig_probes}
\end{center}
\vskip -0.2in
\end{figure*}

We present our validation loss plots in Fig.~\ref{fig_scaling}. We can see a
clear saturating log-log neural scaling law, agreeing with prior work
\citep{ref_cortes1993,ref_kaplan2020,ref_henighan2020}. We find that the saturation
point occurs near our 89 million parameter model, and therefore use our
smallest and largest non-saturated models (AstroPT-1M and AstroPT-89M) to
generate our embedding plots in Fig.~\ref{fig_embeddings}.  We use all our
non-saturated models (that is, \mbox{AstroPT-\{1M, 5M, 12M, 12M, 89M\}}) to
produce our downstream task scaling tests (Fig.~\ref{fig_probes}). 

Figure~\ref{fig_scaling} shows a saturating loss after some log-log scaling,
which we expect is due to the information density in our chosen dataset
\citep{ref_henighan2020}. In future work we will attempt to confirm our
intuition by comparing this result with a LOM trained on other---more
information dense---observational modalities such as galaxy spectra, less
processed galaxy photometry, stellar time series, or mixed modalities.  

To produce an embedding from our AstroPT models we take the averaged outputs of
the models' penultimate layer over the central 64 tokens of a galaxy image.
In Fig.~\ref{fig_embeddings} we show a `Uniform Manifold Approximation and
Projection' \citep[UMAP;][]{ref_mcinnes2018} projection of AstroPT-1M's and
AstroPT-89M's embeddings.  In these projections each hex bin is coloured by
a selected emergent galactic property. We can see a clear structure in each
subplot, which suggests that the model has learnt semantically meaningful
properties of our galaxy dataset. However, it is difficult to conclude from
these projections whether the larger model's embeddings are more informative
than those of the smaller model's. We therefore perform linear probing on
the embedding spaces of our non-saturated models to quantify how physically
informative our model embeddings are, and describe the process in the next
paragraph.

To investigate AstroPT's performance per pretraining FLOP on downstream tasks
we extract embeddings in the same way as described in the previous paragraph,
and use them to train a linear probe mapping from the embedding to a selected
set of emergent physical properties. We train our linear probes on the
validation set (86\,500 galaxies held out during pretraining), and infer on the
test set (86\,500 further held out galaxies). We perform this probing for all
of our models before the loss saturation point that is shown in
Fig.~\ref{fig_scaling}: the 1M, 5M, 12M, 21M, and 89M parameter models. Once we
have the linear probe accuracy of our downstream tasks for each of our tested
models, we run a Spearman's $\rho$ statistical test to measure the correlation
between total pretraining compute spent and downstream task performance.  Our
findings are presented in Fig.~\ref{fig_probes}. We find that there is a clear
positive correlation between downstream task performance and pretraining FLOP
spent.

The results of our linear probe performance per pretraining FLOP are promising, and
we expect this result to carry over to other observational modalities across
domains, neural architectures, and pretraining routines.
Interestingly, we also see `emergent' abilities (or abilities that suddenly
manifest at a certain model scale) here, similarly to what has been shown
in language modelling \citep{ref_wei2022emergent}: the stellar
mass estimation ($M_*$) and tight spiral morphological classification probes see
sudden performance improvements at 12M parameters.

Notably, aside from the properties shown in Fig.~\ref{fig_embeddings} and
Fig.~\ref{fig_probes}, we can to an extent predict the galaxy location and
object ID with a linear probe. This is a result that initially may seem surprising.
However, given that the DESI-LS telescopes used for surveying different
sections of the sky reportedly have distinct effective response curves
\citep{ref_zhou2021}, it is likely that the model captures the
instrumental differences between them.  We can also see this effect in our UMAP
plots in Fig.~\ref{fig_embeddings}, where each `island' corresponds to objects
observed with different DESI-LS telescopes. This effect could be mitigated
downstream if desired via some domain adaptation finetuning, for example via
adversarial domain adaptation \citep{ref_ganin2016,ref_ciprijanovic2020}.

We chose to use linear probing for simplicity---and to robustly show that 
model scale drives downstream task improvement---but of course we can
also use more complicated finetuning methods that would not assume a linear
relationship between our embedding space and the downstream task labels. 
As this study's purpose is to show that self-supervised causal transformer
models can scale in downstream performance in the astronomical domain, we leave
investigation into optimal finetuning methods and therefore comparison to the
state-of-the-art to future work.

\section{Conclusions}
\label{sec_conclusion}

To summarise this paper's technical contributions:
we demonstrated that simple generative autoregressive models can learn
scientifically useful information when pretrained on the surrogate task of
predicting the next $16 \times 16$ pixel patch in a sequence of galaxy image
patches.  Furthermore, we showed that our AstroPT model scales
predictably in downstream task performance as it is pretrained on more compute,
a process that has been shown to be true within natural imagery
\citep[e.g.][]{ref_henighan2020} and natural language \citep[e.g.][]{ref_kaplan2020}. This
is a promising result that suggests that data taken from the observational
sciences would complement data from other domains when used to pretrain a single multimodal
LOM \citep{ref_aghajanyan2023}, and so points towards the use of observational data 
as one solution to the `token crisis' \citep{ref_sevilla2022,ref_xue2023}.

As the AstroPT LOM framework is---by design---very flexible, we expect that similar
autoregressive models can be used across many observational modalities. And
we deliberately designed the AstroPT architecture and generative training
regime to stay as close to the current leading community models as possible. We
took these decisions in the belief that collaborative community development
paves the fastest route towards realising an open source web-scale large observation
model.  We therefore hope that this work seeds further research in this area,
and we release our full dataset, model weights for all trained models
checkpointed across the entire training run, and our training code under the
MIT license to encourage this.
We also welcome direct collaboration---drawing inspiration from EleutherAI's
call to do `science in the open' \citep{ref_phang2022} we developed AstroPT in
public from inception as open-to-all project, and we will continue to build
the AstroPT family of LOMs in public on the UniverseTBD Discord
server\footnote{\url{https://discord.gg/CjMBBJKnFH}}. We warmly invite
potential collaborators to join us.

\section*{Code, model weights, and data availability}

Our code is available on Github here:\\\url{github.com/Smith42/astroPT}\\
Our model weights are available on HuggingFace here:\\\url{hf.co/Smith42/astroPT}\\
The data used to train this model are available here:\\\url{hf.co/datasets/Smith42/galaxies}

\section*{Carbon emissions}

The training of deep learning models requires considerable energy, contributing
to carbon emissions.  To counteract further emission from redundant retraining,
we follow the recommendations of \citet{ref_strubell2019} and make available
our fully trained models and code.  To estimate the carbon equivalent emitted
during this work we used the excellent CodeCarbon (\url{codecarbon.io}), which
estimated a total of {120\,kg\,CO2\,{eq.}} across our final pretraining runs.

\section*{Acknowledgements}

We would like to thank James Geach, Jia-Shu Pan, Kevin Schawinski, Regina
Sarmiento, Mike Walmsley, and Zahra Sharbaf for illuminating
discussions and comments.  MJS would like to thank the Instituto de
Astrof\'isica de Canarias for hosting him while he worked on this project.
This study made use of the Liverpool John Moores University’s Propsero HPC facility (\url{prospero-docs.readthedocs.io}).



\bibliography{main}
\bibliographystyle{icml2024}

\end{document}